\documentclass[final]{IEEEtran}
\pdfoutput=1
\usepackage{amsthm,amssymb,graphicx,multirow,amsmath,color,amsfonts}
\usepackage[caption=false,font=footnotesize]{subfig}
\usepackage[update,prepend]{epstopdf}
\usepackage[noadjust]{cite}
\usepackage[latin1]{inputenc}
\usepackage{tikz}
\usepackage{bbm} 
\usepackage{pdfpages}
\usepackage{multirow}
\usepackage{comment}
\usepackage{algorithm}
\usepackage{algorithmic}
\usepackage{url}
\def\BSTATE{\STATE\hskip-\ALG@thistlm}
\makeatother

\def\nb0{{\mathbf{0}}}
\def\nb1{{\mathbf{1}}}

\newtheorem{lemma}{Lemma}

\newtheorem{remark}{Remark}

\allowdisplaybreaks 

\usepackage{setspace}	

\begin{document}
\graphicspath{{./Figures/}}
\title{
Average Peak Age-of-Information Minimization in UAV-assisted IoT Networks 
}
\author{
Mohamed A. Abd-Elmagid and Harpreet S. Dhillon
\thanks{Copyright (c) 2018 IEEE. Personal use of this material is permitted. However, permission to use this material for any other purposes must be obtained from the IEEE by sending a request to pubs-permissions@ieee.org.

The authors are with Wireless@VT, Department of ECE, Virginia Tech, Blacksburg, VA. Email: \{maelaziz,\ hdhillon\}@vt.edu. This work was supported by the U.S. National Science Foundation under Grant CPS-1739642.
} 
\vspace{-5mm}
}

\maketitle
\begin{abstract}

Motivated by the need to ensure timely delivery of information (e.g., status updates) in the Internet-of-things (IoT) paradigm, this paper investigates the role of an Unmanned aerial vehicle (UAV) as a mobile relay to minimize the average Peak Age-of-information (PAoI) for a source-destination pair. For this setup, we formulate an optimization problem to jointly optimize the UAV's flight trajectory as well as energy and service time allocations for packet transmissions. In order to solve this non-convex problem, we propose an efficient iterative algorithm and establish its convergence analytically. Closed-form solutions for some sub-problems are also provided. One of the sub-problems we solve in this procedure is to jointly optimize the energy and service time allocations for a given trajectory of the UAV. This problem is of interest on its own right because in some cases we may not be able to alter the UAV's trajectory based on the locations of the IoT devices (especially when its primary mission is something else). Our numerical results quantify the gains that can be achieved by additionally optimizing the UAV's trajectory.%

\end{abstract}
\vspace{-3 mm}
\section{Introduction} \label{sec:intro}

A key performance bottleneck in several IoT-enabled applications is the {\em freshness} of the aggregated data measurements of the IoT devices when they reach the destination nodes. For instance, in human safety applications, the out-of-date measurements may result in erroneous controllable decisions, which may be catastrophic. The energy-constrained nature of the IoT devices increases the likelihood of packets loss, which affects the timely delivery of their measurements. This is even more critical for far-off IoT devices whose direct links to the destination nodes may be very poor \cite{7842431}. This necessitates designing new network architectures that efficiently utilize the limited energy at the IoT devices and guarantee timely delivery of their measurements and status updates. Since UAVs are expected to be a key component of future wireless architectures, in this paper we study how they can be naturally utilized to maintain freshness of measurements even if their primary mission is something entirely different.

We employ Age-of-information (AoI) \cite{kaul2012real,yates2012real,kaul2012status,kam2013age,chen2016age} as a metric to quantify the freshness of information at the destination node. In order to put our contribution in context, we first discuss a series of key prior works that introduced and used this concept. The authors of \cite{kaul2012real} characterized the average AoI for a single source-destination pair in which randomly generated measurement packets arrive at the source according to a Poisson process, and then transmitted to the destination using a first-come-first-served (FCFS) discipline. The authors of \cite{kam2013age,yates2012real,kaul2012status} extended the results obtained in \cite{kaul2012real} when the update packets: i) may be delivered out-of-order~\cite{kam2013age}, ii) belong to two independent transmitting sources~\cite{yates2012real}, and iii) are transmitted using a last-come-first-served discipline (LCFS) \cite{kaul2012status}. A common assumption in \cite{kaul2012real,yates2012real,kaul2012status,kam2013age} was that the update packets are always delivered successfully to the destination. Relaxing this, \cite{chen2016age} studied the case in which a packet delivery error may occur probabilistically. The authors of \cite{8006544} studied status update transmissions over unreliable multiaccess channels. Different from these, we consider the scenario in which the quality of the direct link between the source and destination nodes is not good enough to reliably meet the data freshness requirements at the destination node. Due to the flexibility of optimizing the trajectory and energy resource allocation of a UAV, we investigate its role as a mobile relay for PAoI minimization in such cases.

The use of UAVs to improve the performance of wireless networks has attracted considerable attention in the recent past. For instance, \cite{azari2016joint,mozaffari2016efficient,bor2016efficient} study their optimal deployment to maximize network coverage and rate, \cite{chetlur2017downlink} uses tools from stochastic geometry to characterize network coverage, \cite{zeng2016throughput,wu2018joint} characterize optimal flight trajectory of UAVs to maximize network throughput, and \cite{mozaffari2017wireless} used tools from optimal transport theory to characterize the minimum hover time of a UAV. Interested readers are advised to refer to \cite{mozaffari2018tutorial} for a comprehensive survey. Building on these prior efforts, this paper makes the first attempt to efficiently design a UAV's flight trajectory as well as energy and service time allocations in order to minimize PAoI for a source-destination link. 

{\em Contributions.} 
For an IoT-inspired setup in which a UAV acts as a mobile relay for a source-destination pair, we formulate the average PAoI minimization problem to jointly optimize the UAV's flight trajectory, energy allocations and service time durations for transmitting update packets at both the source node and the UAV. For solving this non-convex problem, we propose an efficient iterative algorithm and establish its convergence analytically. As an important sub-problem of this procedure, we characterize the optimal energy and service time allocations for packet transmissions for a given flight trajectory of the UAV. This sub-problem is of particular practical interest in situations where the flight trajectory of the UAV cannot be altered much due to its energy and other mission-specific constraints. Several system design insights are provided both through closed-form characterizations of some sub-problems as well as useful numerical results. 

\section{System Model}

We consider a source-destination pair in which the source node is supposed to send its measurements or status updates to the destination node. As discussed in the previous Section already, we assume that the direct link between the two nodes is weak because of which a UAV is used as a mobile relay. The choice of UAV for this is motivated by its increasing relevance in wireless networks and the fact that it can be used for this purpose even if its secondary mission is something else. In the context of IoT networks, the source node may refer to an IoT gateway (GW) located near IoT devices, which transmits measurements collected from them to the destination node (for instance, a cellular base station) through the assistance of the UAV. We use AoI as the measure of freshness of information at the destination node. The AoI at time $t$ is defined as
\begin{align}\label{Ins_AoI}
a(t) = t - u(t),
\end{align}
where $u(t)$ is the time at which the most recently received packet at the destination node was generated at the source node. For clarity of exposition, we assume that $a(0) = 0$. To characterize the minimum achievable AoI, we assume {\em just-in-time} transmission policy \cite{kaul2012real}. That is, a new update packet is instantaneously generated by the source node and starts its service time right after the current update packet in service is received at the destination node. Therefore, as shown in Fig. 1, $a(t)$ is reset to $d_{i}$ at $t_i =\sum_{j=1}^{i}{d_j}$, $i = 1, \cdots, N$, where $d_i$ denotes the service time of the $i$-th packet and $N$ denotes the total number of packets to be transmitted from the source node to the destination node through the assistance of the UAV. Let $a_{i}$ denote the peak value of $a(t)$ right before it is reset at $t_i$ when the $i+1$-th packet is received at the destination node. Clearly, $a_{i} = d_{i} + d_{i+1},\; i = 1,\cdots, N-1$. For this setup, the average PAoI \footnote{Note that the average PAoI is only a function of service time durations of packet transmissions since we aim at characterizing its minimum achievable value through employing {\em just-in-time} transmission policy. This shows that a similar optimization approach to the one considered in this paper could be used to characterize the optimal transmission policy which minimizes the total delay of packet transmissions.} \cite{costa2016age} is defined as
\begin{align}\label{PAoI}
A_P = \frac{1}{N-1}\sum_{i=1}^{N-1}{d_{i} + d_{i+1}}. 
\end{align}

The UAV is assumed to be equipped with a single antenna, and thus operates in a half-duplex mode. Thus, the service time of the $i$-th packet, $d_{i}$, is partitioned into two phases: i) {\em uplink phase}: during the first portion of the service time, denoted by $d_{i,1}$, the source node transmits the $i$-th packet to the UAV, and ii) {\em downlink phase}: in which the UAV delivers packet $i$ to the destination. We assume that the UAV is flying at a fixed height $h$ with a maximum allowable velocity $V_{\rm max}$. The source and destination nodes are considered to be located on the ground at $L_{S}$ and $L_{D} \in \mathbb{R}^{2}$. Let $q(t)$ denote the projection of the UAV's flight trajectory on the ground at time $t,\; t \in [0,t_{N}]$. For convenience, during the service time of an arbitrary packet $i$, we assume that $q(t) = q_{i,1}$ for $t \in [t_{i-1},t_{i-1}+d_{i,1})$ and $q(t) = q_{i,2}$ for $t \in [t_{i-1}+d_{i,1},t_{i})$, i.e., the projection of the UAV's trajectory on the ground does not change in the uplink or downlink phases with respect to the locations of the source and destination nodes. Thus, the UAV's flight trajectory is approximated by the sequence $\mathbf{Q} = \{q_{1,1},\;q_{1,2},\;q_{2,1},\;q_{2,2},\cdots, q_{N,1},\;q_{N,2}\}$. We study a practical scenario in which the initial and final locations of the UAV mainly depend on its launching and landing locations, and hence are known \cite{zeng2016throughput}. Let $q_0$ and $q_{f}$ denote the horizontal coordinates of the UAV's initial and final locations, respectively, i.e., $q_{1,1} = q_0$ and $q_{N,2} = q_f$. 

Similar to \cite{zeng2016throughput,wu2018joint}, the uplink channels from the source node to the UAV and downlink channels from the UAV to the destination node are assumed to be dominated by the line-of-sight (LoS) links and the Doppler effect due to UAV's mobility is assumed to be well compensated for. Thus, the uplink and downlink channel power gains are given respectively by
\begin{align}\label{up_channel}
g_{i,1} = \nu_0 d_{SU}^{- 2} = \frac{\nu_0}{h^2 + \lVert q_{i,1} - L_{S}\rVert^2},\; i= 1, \cdots, N,
\end{align} 
\begin{align}\label{down_channel}
g_{i,2} = \nu_0 d_{UD}^{- 2} = \frac{\nu_0}{h^2 + \lVert q_{i,2} - L_{D}\rVert^2},\; i= 1, \cdots, N,
\end{align}
where $d_{SU}$ denotes the distance between the source node and the UAV, $d_{UD}$ denotes the distance between the UAV and the destination node, and $\nu_0$ denotes the channel power gain at a reference distance of 1 meter.

Using Shannon's formula, the maximum achievable throughputs in uplink and downlink phases, in bits/Hz, can be expressed respectively as
\begin{align}\label{up_rate}
R_{i,1} = d_{i,1} \log_{2} \left(1 + \dfrac{g_{i,1}E_{i,1}}{\Gamma \sigma^{2} d_{i,1}}\right),\; i= 1, \cdots, N,
\end{align} 
\begin{align}\label{down_rate}
R_{i,2} = d_{i,2} \log_{2} \left(1 + \dfrac{g_{i,2}E_{i,2}}{\Gamma \sigma^{2} d_{i,2}}\right),\; i= 1, \cdots, N,
\end{align}
where $E_{i,1}$ is the energy consumed by the source node to transmit packet $i$ to the UAV, $E_{i,2}$ is the energy consumed by the UAV to deliver packet $i$ to the destination node, $\sigma^{2}$ is the noise power and $\Gamma$ is the signal to noise ratio gap due to a practical modulation and coding scheme used.
\begin{figure}
\centering
\includegraphics[width=0.5\columnwidth]{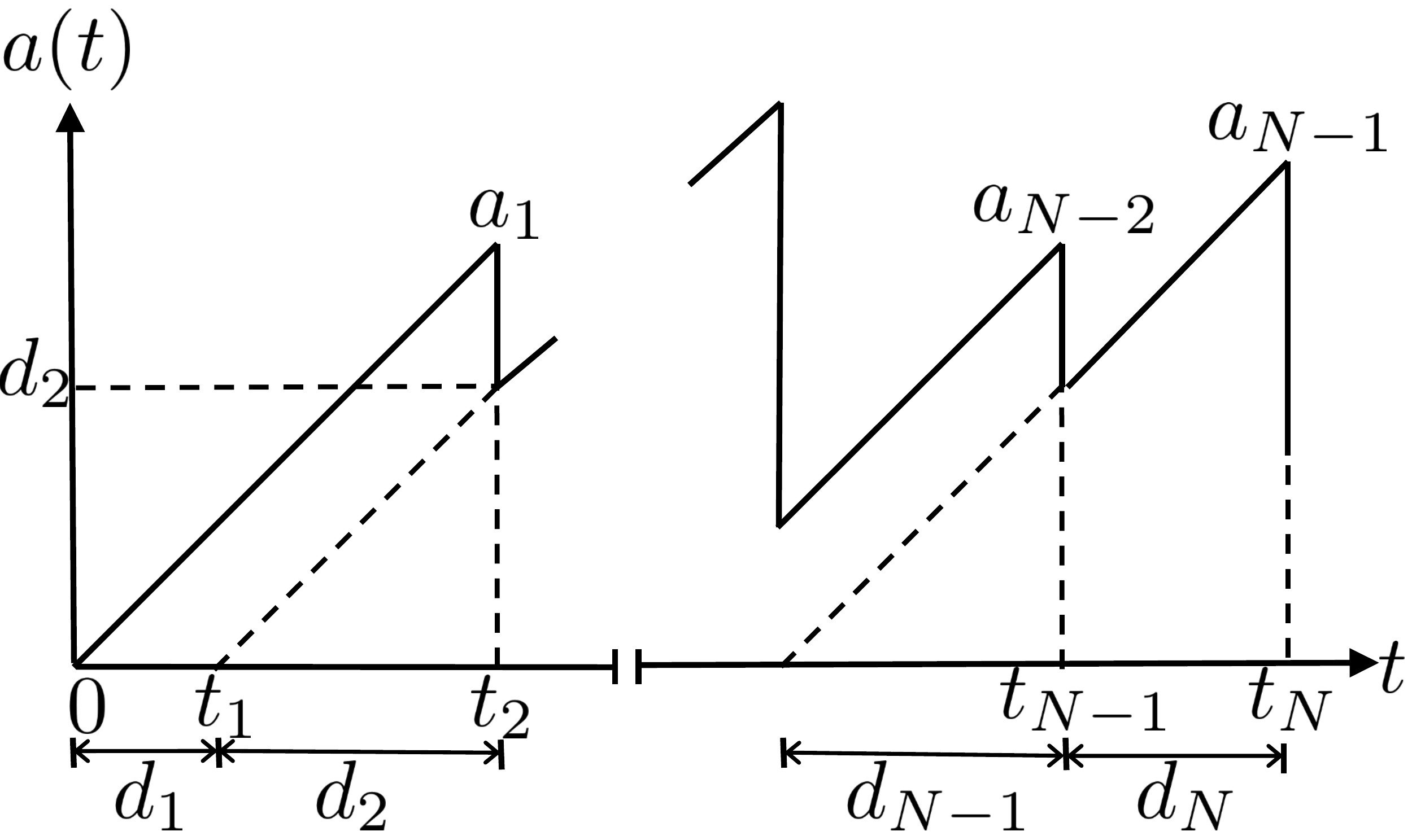}
\caption{AoI evolution vs. time for $N$ update packets.}
\label{fig:illu}
\end{figure}
\section{Problem Formulation and Proposed Solution}
\subsection{Problem Formulation}
Our prime objective is to characterize the minimum average PAoI, as defined in (\ref{PAoI}). Towards this objective we jointly optimize the UAV's trajectory and the energy allocations for transmitting update packets at both the source node and UAV, subject to the UAV's mobility constrains and the total available energy constraints at the source node and UAV. From (\ref{PAoI}) - (\ref{down_channel}), the average PAoI minimization problem can be formulated as  
\begin{align}
\nonumber & \textbf{P1}: \hspace{0.2cm}  && \nonumber\underset{\mathbf{d}_{1},\mathbf{d}_{2},\mathbf{E}_1,\mathbf{E}_2,\mathbf{Q}}{\text{min}}\;\;  \frac{1}{N-1}\sum_{i=1}^{N-1}{d_{i} + d_{i+1}} \\
\label{P1_a}&\text{s.t.} && d_{i,j} \log_{2} \left(1 + \dfrac{g_{i,j}E_{i,j}}{\Gamma \sigma^{2} d_{i,j}}\right) \geq \bar{S},\; \forall j,\; i=1,\cdots, N, \\
\label{P1_b} &&& \sum_{i=1}^{N}{E_{i,1}} \leq E_{S},\;\sum_{i=1}^{N}{E_{i,2}} \leq E_{U},\; \mathbf{E}_1 \succeq \mathbf{0},\; \mathbf{E}_2 \succeq \mathbf{0},\\
\label{P1_c} &&& \lVert q_{i,2} - q_{i,1}\rVert \leq d_{i,1} V_{\rm max},\; i=1,\cdots, N, \\
\label{P1_d} &&& \lVert q_{i+1,1} - q_{i,2}\rVert \leq d_{i,2} V_{\rm max},\; i=1,\cdots, N-1, \\
\label{P1_e} &&&  q_{1,1} = q_{0},\; q_{N,2} = q_{f},
\end{align}
where $E_{S}$ and $E_{U}$ denote the total available energy for transmitting update packets at the source node and UAV, respectively, and $\bar{S} = \frac{S}{B}$ where $B$ denotes the channel bandwidth, and $S$ is the size of each update packet. $\mathbf{E}_{j}=[E_{1,j}, \cdots, E_{N,j}]$ and $\mathbf{d}_{j}=[d_{1,j}, \cdots, d_{N,j}]$ for $j \in \{1,2\}$, $\mathbf{0}$ is a vector of zeros that has the same size as $\mathbf{E}_{j}$ and the symbol $\succeq$ represents the element-wise inequality. The constraints in (\ref{P1_a}) guarantee successful packet transmissions to the UAV in the uplink phase and to the destination node in the downlink phase. The energy constraints at both the source node and UAV are captured in (\ref{P1_b}). Note that we only consider the communication-related energy consumption of the UAV. Under a more general model in which the UAV's propulsion energy is considered, the optimal flight trajectory may reduce to a straight line between the UAV's initial and final locations to minimize the energy consumed by the UAV's flight for long distances. Incorporating the UAV's propulsion energy consumption in \textbf{P1} is left as a promising direction for future work. The constraints in (\ref{P1_c}) and (\ref{P1_d}) model the UAV's mobility constraints. Although the objective function of \textbf{P1} is an affine function in $\left(\mathbf{d}_1,\mathbf{d}_2\right)$, the coupling of different optimization variables $\{d_{i,j},E_{i,j},q_{i,j}\}$ in (\ref{P1_a}) makes \textbf{P1} a non-convex optimization problem. Therefore, \textbf{P1} can not be solved directly using standard convex optimization techniques and characterizing its global optimal solution is very challenging. Employing the block coordinate decent optimization technique, we propose an efficient algorithm for solving \textbf{P1} in the next subsection.

\subsection{Proposed Solution}
Applying the block coordinate decent optimization technique, \textbf{P1} can be solved in an iterative manner as follows. First, for a given UAV's flight trajectory $\mathbf{Q}$, we characterize the optimal energy allocations and service time durations, i.e., $\mathbf{E}_{j}$ and $\mathbf{d}_{j}$ for $j \in \{1,2\}$. Afterwards, the obtained optimal energy and service time allocations are used to update $\mathbf{Q}$ using the successive convex optimization technique. With the updated $\mathbf{Q}$, the procedure is repeated until the value of \textbf{P1}'s objective function converges to a pre-specified accuracy. Specifically, for a given $\mathbf{Q}$, \textbf{P1} reduces to the following optimization problem
\begin{align}
\nonumber & \textbf{P2}: \hspace{0.2cm}  && \nonumber\underset{\mathbf{d}_{1},\mathbf{d}_{2},\mathbf{E}_1,\mathbf{E}_2}{\text{min}}\;\;  \frac{1}{N-1}\sum_{i=1}^{N-1}{d_{i} + d_{i+1}} \\
\label{P2_a}&\text{s.t.} && d_{i,1} \log_{2} \left(1 + \dfrac{\gamma_{i,1}E_{i,1}}{d_{i,1}}\right) \geq \bar{S},\;i=1,\cdots, N, \\
\label{P2_b} &&& d_{i,2} \log_{2} \left(1 + \dfrac{\gamma_{i,2}E_{i,2}}{d_{i,2}}\right) \geq \bar{S},\;i=1,\cdots, N,\\
\nonumber &&& {\rm Eqs.}\; (\ref{P1_b}),\;(\ref{P1_c}),\;(\ref{P1_d}), 
\end{align}
where $\gamma_{i,j} = \frac{g_{i,j}}{\Gamma \sigma^2}$ for $i=1,\cdots, N,$ and $j \in \{1,2\}$. As noted already in the previous Section, although \textbf{P2} is a sub-problem of the generic problem \textbf{P1}, it is important on its own right because in many practical situations, we may not be able to alter the UAV's trajectory based on the IoT device locations, in which case the best we can hope to do is to optimize over the energy allocations and service times (which is \textbf{P2}).
\begin{lemma}\label{lem:1}
\textbf{P2} is a convex optimization problem.
\end{lemma}
\begin{IEEEproof}
First, we note that the objective function of \textbf{P2} is an affine function in $(\mathbf{d}_{1},\mathbf{d}_{2})$. Recall that if $f(x)$ is a concave function, then its perspective function $g(x,t) = t f(\frac{x}{t})$ is also a concave function \cite{optimization_book}. Substituting $t$ with $d_{i,j}$, $R_{i,j}$ is the perspective function of the concave function $\log_{2} \left(1 + \gamma_{i,j}E_{i,j}\right)$ for $i = 1, \cdots, N,$ and $j \in \{1,2\}$. Consequently, $R_{i,j}$ is a concave function and hence the constraints in (\ref{P2_a}) and (\ref{P2_b}) constitute convex sets. In addition, all the remaining constrains are affine in $(\mathbf{d}_{1},\mathbf{d}_{2},\mathbf{E}_1,\mathbf{E}_2)$. This completes the proof.
\end{IEEEproof}
Based on Lemma \ref{lem:1}, \textbf{P2} can be solved using standard convex optimization techniques. Note that the optimal solution of \textbf{P2} can be obtained by solving the following two convex optimization problems in parallel
\begin{align}
\nonumber & \textbf{P2.a}: \hspace{0.5cm}  && \nonumber\underset{\mathbf{d}_{1},\mathbf{E}_1}{\text{min}}\;\;  \frac{1}{N-1}\sum_{i=1}^{N-1}{d_{i,1} + d_{i+1,1}} \\
\nonumber &\text{s.t.} && {\rm Eqs.}\; (\ref{P2_a}),\;(\ref{P1_c}), \\
\nonumber &&& \sum_{i=1}^{N}{E_{i,1}} \leq E_{S},\; \mathbf{E}_{1} \succeq \mathbf{0}. 
\end{align}
\begin{align}
\nonumber & \textbf{P2.b}: \hspace{0.5cm}  && \nonumber\underset{\mathbf{d}_{2},\mathbf{E}_2}{\text{min}}\;\;  \frac{1}{N-1}\sum_{i=1}^{N-1}{d_{i,2} + d_{i+1,2}} \\
\nonumber &\text{s.t.} && {\rm Eqs.}\; (\ref{P2_b}),\;(\ref{P1_d}), \\
\nonumber &&& \sum_{i=1}^{N}{E_{i,2}} \leq E_{U},\; \mathbf{E}_{2} \succeq \mathbf{0}. 
\end{align}
Due to the structural similarity between \textbf{P2.a} and \textbf{P2.b}, in the following we focus on characterizing the optimal solution for \textbf{P2.a}. Define $\theta_i = \frac{\lVert q_{i,2} - q_{i,1}\rVert} {V_{\rm max}}, i=1,\cdots, N,$ and $E_{\rm min} = \sum_{i=1}^{N}{\frac{\theta_i}{\gamma_{i,1}}\left(2^{\bar{S}/\theta_i} - 1\right)}$. When $E_{S} \geq E_{\rm min}$, the following Lemma gives the optimal solution of \textbf{P2.a}.
\begin{lemma}\label{lem:2}
For $E_{S} \geq E_{\rm min}$, the optimal energy and service time allocations of \textbf{P2.a} are given respectively by
\begin{align}\label{opt_energy_case1} 
E_{i,1}^* = \frac{\theta_i}{\gamma_{i,1}}\left(2^{\bar{S}/ \theta_i} - 1\right),\; i=1,\cdots, N,
\end{align}
\begin{align}\label{opt_stime_case1}
d_{i,1}^* = \theta_{i},\; i=1,\cdots, N.
\end{align}
\end{lemma}
\begin{IEEEproof}
Note that $R_{i,1}$ in (\ref{P2_a}) is a monotonically increasing function in both $d_{i,1}$ and $E_{i,1}$. Therefore, from (\ref{P2_a}) and (\ref{P1_c}), the minimum energy required to achieve the minimum allowable service time of packet $i$ is given by (\ref{opt_energy_case1}). Consequently, when $E_{S} \geq E_{\rm min}$, each update packet $i$ could be transmitted with its minimum feasible service time $\theta_i$. This completes the proof.
\end{IEEEproof}
Now, we proceed to characterize the optimal solution of \textbf{P2.a} when $E_{S} < E_{\rm min}$. We begin with the following remark.%
\begin{remark}\label{rem:1}
Note that the constraints in (\ref{P2_a}) should hold with equality at the optimal solution of \textbf{P2.a} (otherwise the objective function could be decreased by either decreasing $d_{i,1}$ for some inactive constraint or increasing the allocated energy for some update packet, with $d_{i,1} > \theta_i$). Furthermore, the total consumed energy limitation constraint should also be satisfied with equality at the optimal solution since otherwise the objective function could be further decreased by allocating more energy for some update packet  with $d_{i,1} > \theta_i$ until the constraint is met with equality. Note that the existence for at least one packet with $d_{i,1} > \theta_i$ is guaranteed by $E_{S} < E_{\rm min}$.
\end{remark}
Taking into account Remark \ref{rem:1}, the energy allocation variables could be substituted from (\ref{P2_a}) into the total consumed energy limitation constraint as $E_{i,1} = \frac{d_{i,1}}{\gamma_{i,1}}\left(2^{\bar{S}/d_{i,1}} - 1\right),\;i=1,\cdots, N$, and removed from the optimization problem. Consequently the Lagrangian of \textbf{P2.a} can be expressed as
\begin{align}\label{lagran}
\mathcal{L}\left(\mathbf{d_1}, \lambda\right) \nonumber &= \frac{1}{N-1}\sum_{i=1}^{N-1}{d_{i,1} + d_{i+1,1}} \\
& + \lambda \left(\sum_{i=1}^{N}{\frac{d_{i,1}}{\gamma_{i,1}}\left(2^{\bar{S}/d_{i,1}} - 1\right)} - E_{S}\right),
\end{align}
where $\lambda$ is the dual variable associated with the energy limitation constraint of the source node. Therefore the dual function, denoted by $\mathcal{G}\left(\lambda\right)$, can be obtained by solving the following optimization problem:
\begin{align}
\nonumber & \textbf{D2.a}: \hspace{0.5cm}  && \nonumber\underset{\mathbf{d}_{1}}{\text{min}}\;\;  \mathcal{L}\left(\mathbf{d_1}, \lambda\right) \\
\nonumber &\text{s.t.} && {\rm Eq.}\;(\ref{P1_c}). 
\end{align}
Hence, the dual problem will be: $\underset{\lambda}{\text{max}} \; \mathcal{G}\left(\lambda\right)$. The following Lemma gives the optimal solution of \textbf{D2.a}.
\begin{lemma}\label{lem:3}
Given $\lambda$, the optimal service time allocations of \textbf{D2.a} are given by
\begin{align}\label{opt_stime}
d_{i,1}^* = {\rm max}\{\theta_i,\frac{\bar{S}}{x_i^*}\},\; i=1,\cdots, N,
\end{align}
where $x_i^* > 0$ is the unique solution of $f(x_i) = \psi_i$, where $f(x_i)$ and $\psi_i$ are given respectively by
\begin{align}\label{f_x_i}
f(x_i) = 2^{x_i}\left(\ln(2)x_i - 1\right) + 1,\; i=1,\cdots, N,
\end{align}
\begin{align}\label{psi_i}
\psi_i = 
\begin{cases}
\frac{\gamma_{i,1}}{\lambda(N-1)},\; i \in \{1,N\},\\
\frac{2\gamma_{i,1}}{\lambda(N-1)},\; i = 2, \cdots, N-1.
\end{cases}
\end{align}
\end{lemma}
\begin{IEEEproof}
It can be easily shown that there exists $\mathbf{d}_{1}$ that strictly satisfies (\ref{P1_c}). Consequently, according to Slater's condition~\cite{optimization_book}, strong duality holds for this problem; therefore, the KKT conditions given below are necessary and sufficient for the global optimality of \textbf{D2.a}:~
\begin{align}\label{KTT}
    \frac{\partial}{\partial d_{i,1}} \mathcal{L} = c_{i} + \frac{\lambda \left(N -1\right)}{\gamma_{i,1}}\left[2^{\bar{S} / d_{i,1}}\left(1 - \frac{\ln(2)\bar{S}}{d_{i,1}}\right) - 1 \right] = 0,
\end{align}
where $c_{i} = 1$ for $i \in\{1,N\}$ and $c_{i} = 2$ otherwise. Defining the variable $x_{i} = \frac{\bar{S}}{d_{i,1}}$, (\ref{KTT})~can be reformulated as $f(x_i) = \psi_i$, where $f(x_i)$ and $\psi_i$ are given respectively in (\ref{f_x_i}) and (\ref{psi_i}). Since $\frac{\rm d}{{\rm d} d_{i,1}}f(x_i) = (\ln(2))^2 x 2^x $, $f(x_i)$ is a monotonically increasing function of $x_i \geq 0$, with $f(0) = 0$. Therefore, there exists a unique solution $x_i^* > 0$ that satisfies $f(x_i^*) = \psi_{i}$ with $\lambda^* > 0$, and hence $\{d_{i,1}^*\}$ are given as in (\ref{opt_stime}).
\end{IEEEproof}
\begin{remark}\label{rem:2}
Note that $\psi_i$ is directly proportional to $\gamma_{i,1}$, and hence $x_i^*$ is also directly proportional to $\gamma_{i,1}$. Therefore, Lemma \ref{lem:3} demonstrates the fact that the closer the UAV is to the source node, the lower is the achievable service time for update packet transmission from the source node. Furthermore, we observe that the minimum allowable packet service time is restricted by $\theta_i$, which is imposed by the maximum velocity of the UAV.
\end{remark}

\begin{algorithm}[t!]
\caption{\textbf{P1} solver.}\label{algo}
\begin{algorithmic}
 \STATE 1. Initialize: $n = 0$.
 \STATE 2. Repeat
 \STATE \hspace{1cm} (1) Compute $\mathbf{E_1}^{n},\;\mathbf{E_2}^{n},\;\mathbf{d_1}^{n}$ and $\mathbf{d_2}^{n}$ by solving \textbf{P2} with given $\mathbf{Q}^{n}$.
 \STATE \hspace{0.9cm} (2) Compute $\mathbf{Q}^{n+1}$ by solving \textbf{P3} using the successive convex optimization technique with given $\mathbf{E_1}^{n},\;\mathbf{E_2}^{n},\;\mathbf{d_1}^{n}$ and $\mathbf{d_2}^{n}$.
 \STATE \hspace{1cm} (3) $n \leftarrow n + 1$.
 \STATE 3. Until the fractional decrease of $A_P$ is below $\epsilon$.
\end{algorithmic}
\end{algorithm} 
We now summarize the procedure for solving \textbf{P2.a}, which in turn characterizes the optimal solution for \textbf{P2} (recall that the structure of \textbf{P2.b} is similar to that of \textbf{P2.a}). If $E_{S} \geq E_{\rm min}$, Lemma \ref{lem:2} can be used to obtain the optimal solution of \textbf{P2.a}. Otherwise, for a given $\lambda > 0$, the optimal service time and energy allocations are evaluated from Lemma \ref{lem:3}. Afterwards, $\lambda$ is updated using the sub-gradient method with the sub-gradient of $G(\lambda)$, given by $\nu$, where
\begin{align}\label{sub_grad}
\nu = \sum_{i=1}^{N}{\frac{d_{i,1}^*}{\gamma_{i,1}}\left(2^{\bar{S}/d_{i,1}^*} - 1\right)} - E_{S} ,
\end{align}
then the updated $\lambda$ is used to evaluate the optimal service time and energy allocations from Lemma \ref{lem:3} again and the procedure is repeated until the stopping criteria of the sub-gradient method is met. Consequently, the last updated $\lambda$ will be the optimal solution of the dual problem, and the corresponding service time and energy allocations will be the optimal solution of \textbf{P2.a}. 

Next, for a given energy and service time allocations, \textbf{P1} becomes a feasibility optimization problem in $\mathbf{Q}$. Thus, our goal is to investigate the feasible trajectory $\mathbf{Q}$ which maximizes the minimum achievable throughput over all the transmissions of update packets, i.e., $\underset{i}{\text{min}}\{R_{i,1},R_{i,2}\}$, so that the objective function of \textbf{P2} is guaranteed to decrease in the next iteration via reducing the service time durations. Thus, the optimal $\mathbf{Q}$ can be obtained by solving the following optimization problem
\begin{align}
\nonumber & \textbf{P3}: \hspace{0.5cm}  && \nonumber\underset{\bar{R},\mathbf{Q}}{\text{max}}\;\;  \bar{R} \\
\label{P3_a}&\text{s.t.} && R_{i,1} \geq \bar{R},\; R_{i,2} \geq \bar{R},\;i=1,\cdots, N, \\
\nonumber  &&& {\rm Eqs.}\; (\ref{P1_c}),\;(\ref{P1_d}),\;(\ref{P1_e}), 
\end{align}
where $\bar{R}$ is an auxiliary variable which denotes the minimum achievable throughput over all transmissions of update packets. Clearly, the non-convex constraints in (\ref{P3_a}) makes \textbf{P3} a non-convex optimization problem. However, the problem's structure could be exploited to obtain a lower bound on its optimal solution using the successive convex optimization technique which is guaranteed to converge to at least a local optimal solution of \textbf{P3}. At each iteration of the successive convex optimization technique, a lower bound on \textbf{P3} is maximized. Note that, for a given $\{E_{i,j},d_{i,j}\}$, $R_{i,j}$ is a convex function with respect to $\lVert q_{i,j} - L_{j}\rVert^2$, where $L_{1} = L_{S}$ and $L_{2} = L_{D}$. Therefore, $R_{i,j}$ could be globally lower bounded by its first-order Taylor expansion at any point $q_{i,j}$. Using the lower bounded functions on $\{R_{i,j}\}$, denoted by $\{R_{i,j}^{\rm lb}\}$, defined at an initial trajectory $\mathbf{Q}$, \textbf{P3} could be solved iterativly until $\bar{R}$ converges to a pre-specified accuracy. In the $n$-th interation, $R_{i,j}^{\rm lb}$ is given by
\begin{align}
R_{i,j}^{\rm lb} =C_1^{n} - C_2^{n} \left(\lVert q_{i,j} - L_{j}\rVert^2 - \lVert q_{i,j}^{n} - L_{j}\rVert^2\right),
\end{align}
where 
\begin{align}
C_{1}^{n} = d_{i,j} \log_{2} \left(1 + \dfrac{\nu_{0}E_{i,j}}{\Gamma \sigma^{2} C_3^{n} d_{i,j}}\right),
\end{align}
\begin{align}
C_{2}^{n} = \dfrac{ \nu_0 \log_{2}({\rm e}) d_{i,j} E_{i,j}}{C_3^{n} \left(\Gamma \sigma^2 C_3^{n} d_{i,j} + \nu_0 E_{i,j}\right)},
\end{align}
\begin{align}
C_{3}^{n} = h^2 + \lVert q_{i,j}^{n} - L_{j}\rVert^2,
\end{align}
where $\mathbf{Q}^{n}$ is the optimal trajectory in the previous iteration $n - 1$. Note that $R_{i,j}^{\rm lb}$ is a concave quadratic function in $q_{i,j}$, and hence at each iteration \textbf{P3} becomes a convex quadratically constrained quadratic program (QCQP), which can be solved efficiently by standard convex optimization solvers, e.g., CVX.

Finally, \textbf{P1}'s solution could be obtained by solving \textbf{P2} and \textbf{P3} iterativly until the average PAoI converges to a pre-determined accuracy. The solution of \textbf{P1} is summarized in Algorithm \ref{algo}. Furthermore, the convergence of Algorithm \ref{algo} is established in the following remark.
\begin{remark}\label{rem:3}
Recall that \textbf{P1} is a feasibility optimization problem in $\mathbf{Q}$ for a given energy and service time allocations. Thus, our objective was to characterize optimal $\mathbf{Q}$ which maximizes the minimum achievable throughput over all transmissions of update packets via solving \textbf{P3}. This would certainly guarantee that the objective value of \textbf{P2} (average PAoI) is non-increasing across different iterations. Since the average PAoI is bounded from below, Algorithm \ref{algo} is guaranteed to converge.
\end{remark}
\begin{figure}[t!]
\centering
\includegraphics[width=0.8\columnwidth]{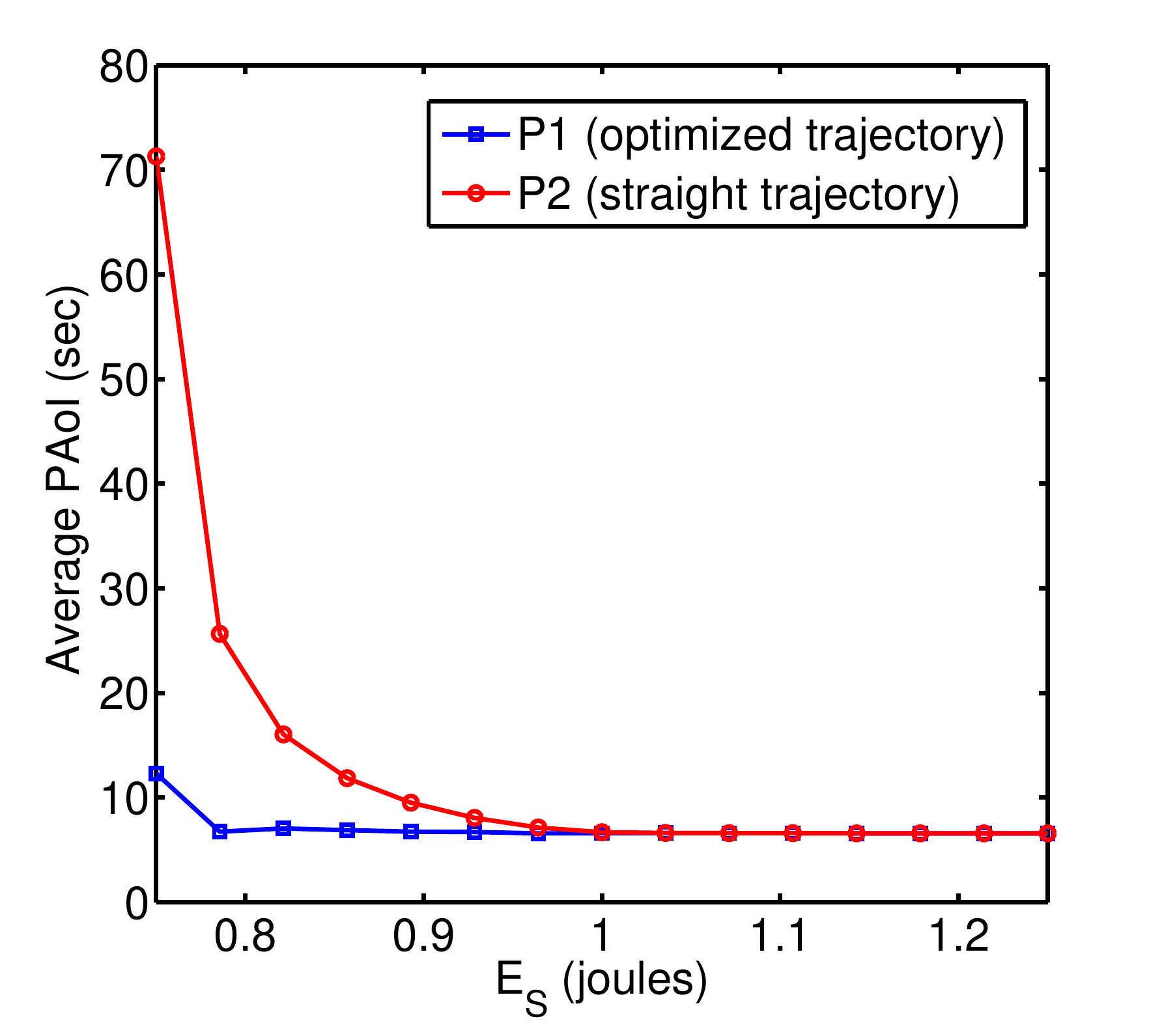}
\caption{Impact of $E_s$ on average PAoI.}
\label{fig:1}
\end{figure}
\begin{figure}
\centering
\includegraphics[width=0.8\columnwidth]{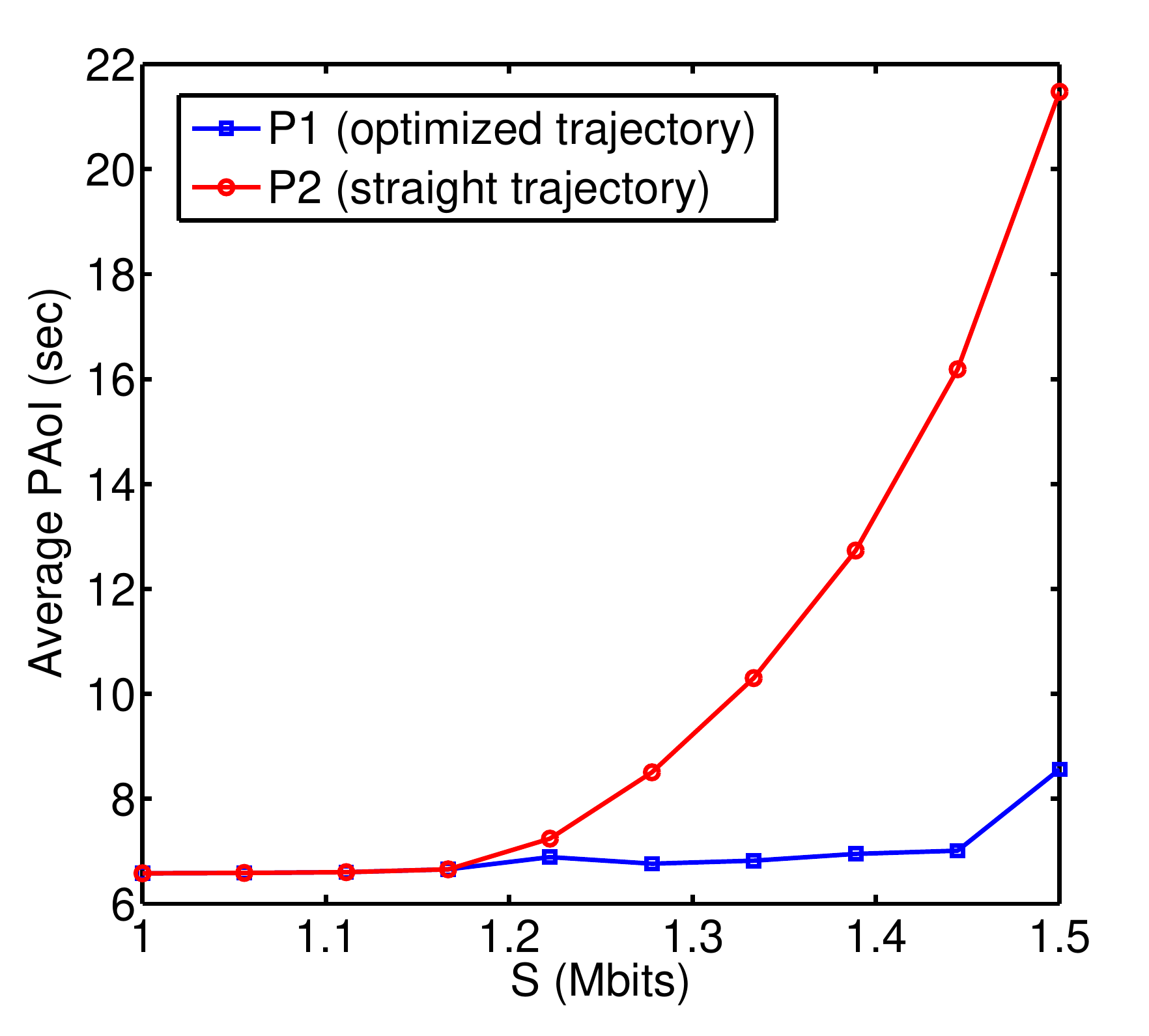}
\caption{Impact of $S$ on average PAoI.}
\label{fig:2}
\end{figure}  
\section{Numerical Results}
In this section, we provide numerical results showing the impact of system parameters on our proposed design. The locations of the source and destination nodes are given respectively by $L_{S} = [-800,\;800]^{\rm T}$ and $L_{D} = [800,\; 800]^{\rm T}$. The UAV is assumed to fly at a fixed height $h = 100$ meters with a maximum allowable velocity $V_{max} = 50$ meters/sec starting from $q_0 = [-800,\;0]^{\rm T}$ to $q_f = [800,\;0]^{\rm T}$. Unless otherwise specified, we consider the following parameters: $N = 10$, $S = 1$ Mbits, $B = 1$ MHz, $E_{S} = E_{U} = 1.25$ joules, $\nu_0 = - 47$ dB, $\Gamma = 10$ dB and $\sigma^2 = - 100$ dBm. 

In Figs. \ref{fig:1} and \ref{fig:2}, we examine the impact of the UAV's trajectory design on the minimum achievable average PAoI. Towards this objective, we quantify the performance gains of the optimized trajectory (obtained from solving \textbf{P1}) over the simple straight trajectory, denoted by $\mathbf{Q}^{\rm s}$. Clearly, in the latter scenario, the minimum achievable average PAoI is obtained from solving \textbf{P2} with a given trajectory $\mathbf{Q}^{\rm s}$, where $q_{i,1}^{\rm s} = [- 800 + (2 i - 2) \frac{1600}{2 N - 1},\;0]^{\rm T} $ and $q_{i,2}^{\rm s} = [- 800 + (2 i - 1) \frac{1600}{2 N - 1},\;0]^{\rm T}$ for $i = 1,\cdots, N$. Note that $\mathbf{Q}^{0}$ is set to be $\mathbf{Q}^{\rm s}$ in Algorithm \ref{algo}. The key message is that the optimal design of the UAV's flight trajectory is of significant importance to ensure the freshness of collected status updates at the destination, especially when the available energy at the source node and UAV is limited (Fig.\ref{fig:1}) or the size of the update packet is large (Fig.\ref{fig:2}), i.e., more energy is needed to deliver each packet to the destination node. Furthermore, it is observed that the minimum average PAoI saturates when $E_s$ exceeds a certain energy threshold or $S$ becomes less than some update packet size. This is due to the fact that there are no energy limitations at the source node and UAV in such cases. The resulting saturated value can be evaluated from Lemma \ref{lem:2} with a given flight trajectory $\mathbf{Q}^{\rm s}$.

\begin{figure}[t!]
\centering
\includegraphics[width=0.8\columnwidth]{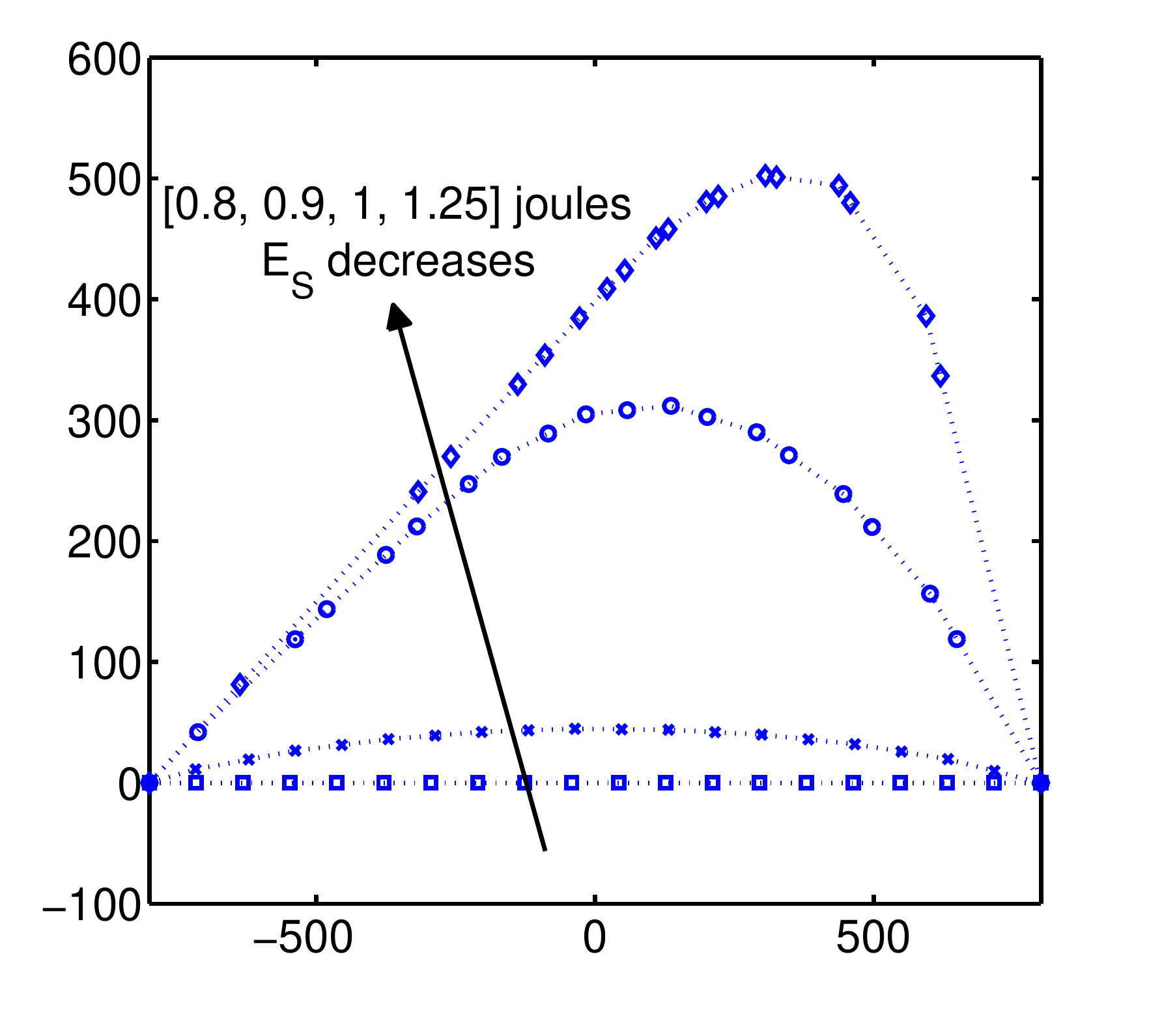}%
\caption{Impact of $E_s$ on optimized trajectory.}
\label{fig:3}
\end{figure}

\begin{figure}
\centering
\includegraphics[width=0.8\columnwidth]{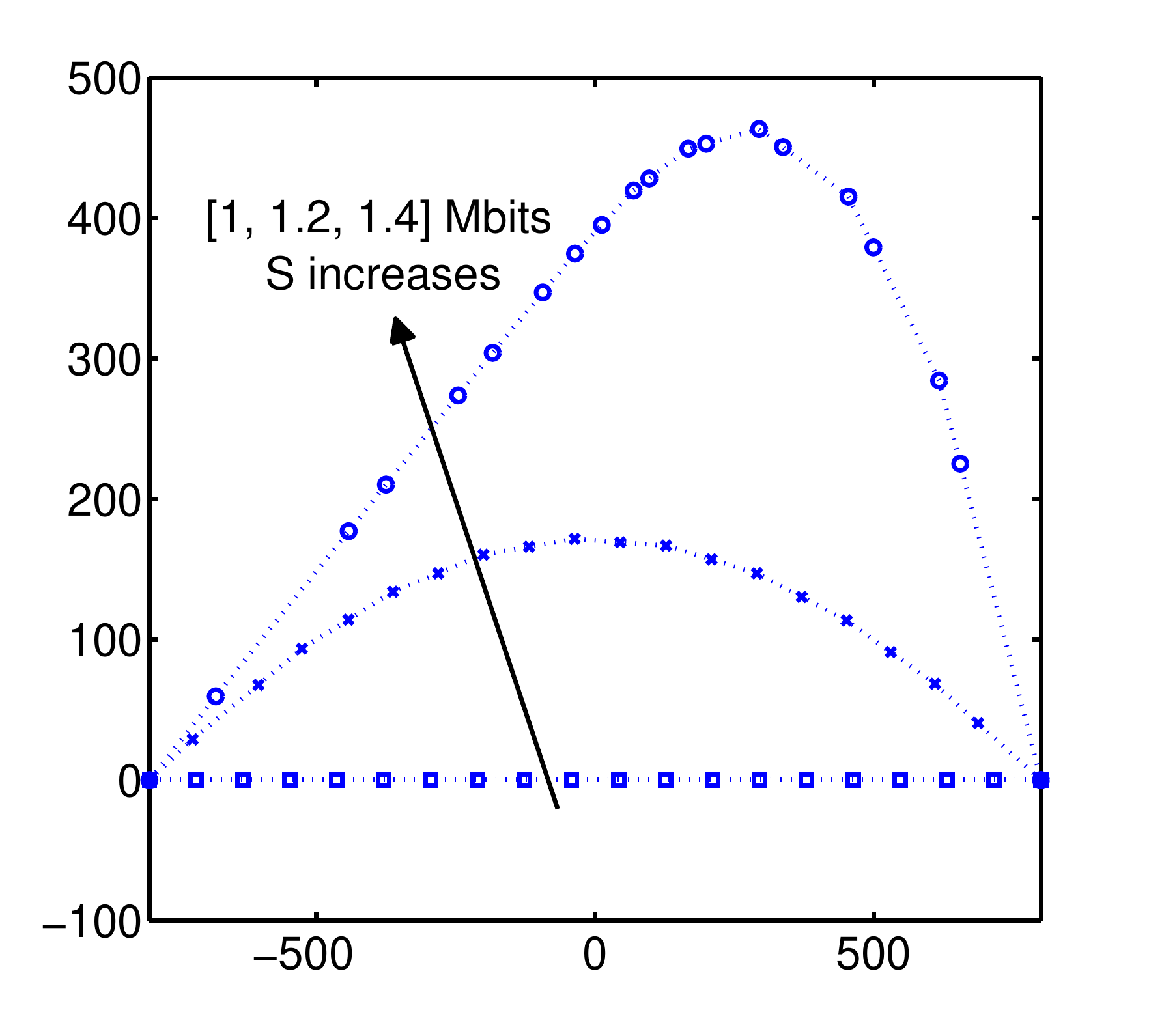}%
\caption{Impact of $S$ on optimized trajectory.}
\label{fig:4}
\end{figure}

In Figs. \ref{fig:3} and \ref{fig:4}, we explore the nature of the UAV's optimal flight trajectory obtained from solving \textbf{P1}. Clearly, the flying distance between each two consecutive phases and its direction mainly depend on the amount of energy available at the source node and the UAV, the size of update packet and the maximum allowable UAV's flying velocity. Particularly, as $E_{S}$ decreases and/or $S$ increases, the UAV moves as close as possible to the source node (destination node) in the downlink phase (uplink phase) to achieve the minimum average PAoI.
\section{Conclusion} 

In this paper, we studied the problem of PAoI minimization for an IoT-inspired setup in which a source node is supposed to transmit its measurements to a destination node through a UAV that acts as a mobile relay. For this setup, we formulated an optimization problem to jointly optimize the UAV's trajectory, as well as energy and service time allocations in order to minimize the overall PAoI. Even though this problem is non-convex, we showed that it can be efficiently solved using our proposed iterative algorithm, which is guaranteed to converge. 

Multiple key design insights were drawn from our results. For instance, they revealed that the optimal design of the UAV's flight trajectory achieves significant performance gains especially when the available energy at the source node and UAV is limited and/or when the size of the update packet is large. Furthermore, they showed the impact of system design parameters on the nature of the UAV's optimal flight trajectory. Finally, they quantified the available energy requirements at both the UAV and source node to achieve the minimum average PAoI. 

This work has many possible extensions. For instance, in this paper, we considered only the communication-related energy consumption of the UAV. One possible extension is to incorporate the UAV's propulsion energy in the problem formulation as well. Another promising avenue of future work is to develop an online algorithm for UAV's flight trajectory optimization.


\bibliographystyle{IEEEtran}
\bibliography{Manuscript}

\begin{thebibliography}{10}
\providecommand{\url}[1]{#1}
\csname url@samestyle\endcsname
\providecommand{\newblock}{\relax}
\providecommand{\bibinfo}[2]{#2}
\providecommand{\BIBentrySTDinterwordspacing}{\spaceskip=0pt\relax}
\providecommand{\BIBentryALTinterwordstretchfactor}{4}
\providecommand{\BIBentryALTinterwordspacing}{\spaceskip=\fontdimen2\font plus
\BIBentryALTinterwordstretchfactor\fontdimen3\font minus
  \fontdimen4\font\relax}
\providecommand{\BIBforeignlanguage}[2]{{%
\expandafter\ifx\csname l@#1\endcsname\relax
\typeout{** WARNING: IEEEtran.bst: No hyphenation pattern has been}%
\typeout{** loaded for the language `#1'. Using the pattern for}%
\typeout{** the default language instead.}%
\else
\language=\csname l@#1\endcsname
\fi
#2}}
\providecommand{\BIBdecl}{\relax}
\BIBdecl

\bibitem{7842431}
H.~S. Dhillon, H.~Huang, and H.~Viswanathan, ``Wide-area wireless communication
  challenges for the {Internet} of {Things},'' \emph{IEEE Commun. Mag.},
  vol.~55, no.~2, pp. 168--174, Feb. 2017.

\bibitem{kaul2012real}
S.~Kaul, R.~Yates, and M.~Gruteser, ``Real-time status: How often should one
  update?'' in \emph{Proc., IEEE INFOCOM}, 2012.

\bibitem{yates2012real}
R.~D. Yates and S.~Kaul, ``Real-time status updating: Multiple sources,'' in
  \emph{Proc., IEEE ISIT}, 2012.

\bibitem{kaul2012status}
S.~K. Kaul, R.~D. Yates, and M.~Gruteser, ``Status updates through queues,'' in
  \emph{Proc., IEEE CISS}, 2012.

\bibitem{kam2013age}
C.~Kam, S.~Kompella, and A.~Ephremides, ``Age of information under random
  updates,'' in \emph{Proc., IEEE ISIT}, 2013.

\bibitem{chen2016age}
K.~Chen and L.~Huang, ``Age-of-information in the presence of error,'' in
  \emph{Proc., IEEE ISIT}, 2016.

\bibitem{8006544}
R.~D. Yates and S.~K. Kaul, ``Status updates over unreliable multiaccess
  channels,'' in \emph{Proc., IEEE ISIT}, 2017.

\bibitem{azari2016joint}
M.~M. Azari, F.~Rosas, K.-C. Chen, and S.~Pollin, ``Joint sum-rate and power
  gain analysis of an aerial base station,'' in \emph{Proc., IEEE Globecom
  Workshops}, 2016.

\bibitem{mozaffari2016efficient}
M.~Mozaffari, W.~Saad, M.~Bennis, and M.~Debbah, ``Efficient deployment of
  multiple unmanned aerial vehicles for optimal wireless coverage,'' \emph{IEEE
  Commun. Letters}, vol.~20, no.~8, pp. 1647--1650, 2016.

\bibitem{bor2016efficient}
R.~I. Bor-Yaliniz, A.~El-Keyi, and H.~Yanikomeroglu, ``Efficient 3-d placement
  of an aerial base station in next generation cellular networks,'' in
  \emph{Proc., IEEE ICC}, 2016.

\bibitem{chetlur2017downlink}
V.~V. Chetlur and H.~S. Dhillon, ``Downlink coverage analysis for a finite 3{D}
  wireless network of unmanned aerial vehicles,'' \emph{IEEE Trans. on
  Commun.}, vol.~65, no.~10, pp. 4543--4558, 2017.

\bibitem{zeng2016throughput}
Y.~Zeng, R.~Zhang, and T.~J. Lim, ``Throughput maximization for uav-enabled
  mobile relaying systems,'' \emph{IEEE Trans. on Commun.}, vol.~64, no.~12,
  pp. 4983--4996, 2016.

\bibitem{wu2018joint}
Q.~Wu, Y.~Zeng, and R.~Zhang, ``Joint trajectory and communication design for
  multi-uav enabled wireless networks,'' \emph{IEEE Trans. on Wireless
  Commun.}, vol.~17, no.~3, pp. 2109--2121, 2018.

\bibitem{mozaffari2017wireless}
M.~Mozaffari, W.~Saad, M.~Bennis, and M.~Debbah, ``Wireless communication using
  unmanned aerial vehicles {(UAVs)}: Optimal transport theory for hover time
  optimization,'' \emph{IEEE Trans. on Wireless Commun.}, vol.~16, no.~12, pp.
  8052--8066, 2017.

\bibitem{mozaffari2018tutorial}
M.~Mozaffari, W.~Saad, M.~Bennis, Y.-H. Nam, and M.~Debbah, ``A tutorial on
  uavs for wireless networks: Applications, challenges, and open problems,''
  2018, available online: arxiv.org/abs/1803.00680.

\bibitem{costa2016age}
M.~Costa, M.~Codreanu, and A.~Ephremides, ``On the age of information in status
  update systems with packet management,'' \emph{IEEE Trans. on Info. Theory},
  vol.~62, no.~4, pp. 1897--1910, 2016.

\bibitem{optimization_book}
S.~Boyd and L.~Vandenberghe, \emph{Convex optimization}.\hskip 1em plus 0.5em
  minus 0.4em\relax Cambridge university press, 2004.

\end{thebibliography}

\end{document}